\documentclass[pra,twocolumn]{revtex4}
\usepackage{graphicx}
\usepackage{graphics}
\usepackage{amssymb}
\usepackage{epstopdf}
\usepackage{color}
\usepackage{subfigure}

\renewcommand{\d}{{\textrm{d}}}

\newcommand{\tr}{{\textrm{tr}}}
\newcommand{\<}{\langle}
\renewcommand{\>}{\rangle}

\begin{document}

\title{Spatial Entanglement From Off-Diagonal Long Range Order in a BEC }
\author{Libby Heaney$^1$, Janet Anders$^2$, Dagomir Kaszlikowski$^2$ and Vlatko Vedral$^{1,2}$}
\affiliation{The School of Physics and Astronomy, University of Leeds, Leeds LS2 9JT, UK$^1$.\\
Department of Physics, National University of Singapore, 117542 Singapore, Singapore$^2$}
\begin{abstract}
We investigate spatial entanglement - particle-number entanglement between regions of space - in an ideal Bosonic gas.  We quantify the amount spatial entanglement around the transition temperature for condensation ($T_{C}$) by probing the gas with two localised two-level systems.  We show that spatial entanglement in the gas is directly related to filling of the ground state energy level and therefore to the off diagonal long range order (ODLRO) of the system and the onset of condensation.  
\end{abstract}
\maketitle
 
 \section{Introduction}
 
Entanglement has been the subject of intense investigation since it was realised that its presence in quantum physics allows for new fast algorithms in computation \cite{NielsenChuang}.  For this, coherent control over a large number of qubits is needed, so it makes sense to study systems that already possess an amount of natural entanglement, for instance spin systems \cite{Toth05} and between spatial regions in a Bose  gas \cite{Anders06, Heaney06}, then one can be sure that under certain conditions entanglement exists even if the system is open.   However, the entanglement found in such systems is more than just a computational resource, it is related to phase transitions in spin chains \cite{Osterloh05, Osborne02, Vidal03} and to superconductivity and superfluidity \cite{Brandao05}. 

Normally, many-body entanglement in such systems is detected, sometimes using macroscopic variables, such as temperature \cite{Vedral04}, magnetic susceptibility \cite{Vedral03} and heat capacity \cite{Wiesniak05a}, that witness the entanglement.  One can then extract the entanglement for other purposes
\cite{Chiara06} using probe systems.

In this paper, unlike these previous approaches, we use two localised probe systems to {\it measure} the entanglement between spatial regions in a Bose gas.  We refer to this type of entanglement as spatial entanglement, which manifests itself as non-local particle number correlations between different regions of space.  Therefore, by locally probing these regions with effective qubits, we skip the need to find an entanglement witness.  We can investigate entanglement without introducing probes \cite{Heaney06}, but in order to make our procedure fully operational we want to show that the entanglement is not an artefact of mathematics. On qubit-type systems standard experiments for the existence of entanglement can be run thereby simplifying the discussion of entanglement in a continuous variable (CV) system.  

In this paper we will show explicitly that spatial entanglement only exists between two distant localised volumes in a Bosonic gas below the critical temperature for a BEC ($T_C$).  Thus the existence of spatial entanglement is directly related to the off-diagonal long range order (ODLRO) of the system.  In fact it is the order in the Bose gas below $T_C$ that allows spatial entanglement between two distant regions to form.

In the remaining parts of the introduction we will discuss the concept of entanglement between regions of space in \ref{sec:EntSpatModes} and discuss the notion of ODLRO in \ref{sec:ODLRO}.  The rest of the paper will be dedicated to calculating the spatial entanglement contained in a Bose gas.  Firstly we discuss the interaction of the probes and the gas in \ref{sec:interaction}, followed by calculation of the amount of entanglement between the probes in sections \ref{sec:entofprobes} and \ref{sec:negativity}.  Finally the results will be discussed in section \ref{sec:resultsanddiscussion} and in the conclusions in \ref{sec:conclusions}.

\subsection{Entanglement between spatial modes}\label{sec:EntSpatModes}
\begin{figure}[t]\centering
{\includegraphics[width=0.4\textwidth]{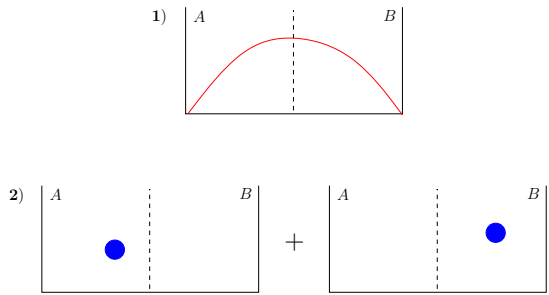}
\caption{A simple example illustrating why spatial entanglement must be considered in second quantisation.  Both diagrams show a particle in the lowest energy level of an infinite square well.  The well has been split into two halves, not physically, but conceptually as this allows one to speak about regions A and B.  Diagram 1 illustrates the situation in first quantisation.  In this representation the particle occupies the wavefunction as shown and if one wants to speak about the state of the particle w.r.t. regions A and B, one can say that the particle is in a superposition of A and B.  However, one cannot speak about entanglement of A and B as the Hilbert space of the whole system is a direct sum of the Hilbert spaces for A and B and there are no observables that we can be measured in regions A or B alone.  Diagram 2 shows the same situation but described in the language of second quantisation.  In this case the regions A and B become the spatial modes and the particle now occupies both the regions non-locally where the state of the two regions would be $|\psi_{AB}\rangle\propto|1\rangle_A\otimes|0\rangle_B+|0\rangle_A\otimes|1\rangle_B$.   In this representation the Hilbert space is a tenosr product of Hilbert spaces of regions A and B and we can now measure the number of particles in each region.  The particle that was represented by a matter wave in diagram 1 is still found across the entire well, but the superposition of left and right in first quantisation becomes particle number entanglement between the regions in second quantisation.    }\label{fig:Fig1}}
\end{figure}

Let us take a closer look at spatial entanglement itself.  When investigating entanglement between spatial regions in a CV system, one must work in second quantisation, where one speaks about the occupation number of modes, because to correctly define entanglement the Hilbert space of a given system must have a tensor product structure. In the first quantisation the Hilbert space of two spatial regions is a direct sum of each region and therefore does not have the required tensor product structure.   However, in this paper we use second quantisation where two regions of space are the modes which are occupied by a given number of particles (excitations).  The entanglement therefore takes the form of non-local particle number correlations between the regions and not between different Bosons in the gas, or their internal degrees of freedom.  
This is because even non-interacting gases in first quantisation description do have coupling between different regions in the second quantisation formulation and this is what in fact leads to the spatial entanglement.    A toy model is given in Fig.~\ref{fig:Fig1} that helps make clear the idea of spatial entanglement as it will be central to the rest of this paper.   

In this paper we use two localised probes coupled for a short time to two distant regions of a Bose gas to measure the spatial entanglement.  To measure the entanglement correctly we must ensure that the probes act with only local operations on the spatial modes.  
The probes cannot couple to the Bosons own degrees of freedom, like momentum, as their eigenfunctions do not form a local basis in the artificially-defined regions but are meaningful only with respect to the entire well.  However, the probes can couple locally to the number of Bosons in each region, even though the regions are populated non-locally, as in this case the number of Bosons in the spatial modes form complete bases for the modes.  To summarise: between the probes we assume no direct interaction and any non-local coupling between them must originate in the non-local nature of the gas they are coupled to.Ê

\subsection{Off-diagonal long-range order (ODLRO) in a Bose gas}\label{sec:ODLRO}

In a BEC ODLRO \cite{Penrose56, Yang62} is present in its simplest form \cite{LRO} when the one-body reduced density matrix 
\begin{equation}
\rho_1(\vec{r},\vec{r}^{\, \prime})=\langle\Psi^{\dagger}(\vec{r})\Psi(\vec{r}^{\, \prime})\rangle
\end{equation}
 is finite as $|\vec{r}-\vec{r}^{\, \prime}|\rightarrow\infty$, where $\Psi^{\dagger}(\vec{r})$ and $\Psi(\vec{r})$ are the creation and annihilation operators of particles at point $\vec{r}$.  The existence of ODLRO testifies that two distance points of a BEC have become strongly correlated and a new thermodynamic phase emerges, the condensed phase.    As the one-body density matrix can be expressed in diagonal form
 \begin{equation}
 \rho_1(\vec{r},\vec{r}^{\, \prime})=\sum_{\vec k} \, N_{\vec k} \, \phi_{\vec k}(\vec{r}) \, \phi^*_{\vec k} (\vec{r}^{\, \prime})
 \end{equation}
one can connect the ODLRO with the eigenvalues $N_{\vec k}$ - the number of particles occupying the single-particle states $\phi_{\vec k}$ of a system.  For instance a BEC occurs when the lowest of the single particle states $\phi_0$ is occupied in macroscopic way, i.e. $N_0/N$ is finite, where $N$ is the total number of particles.  We shall see that it is the enormous occupation of the ground state mode that accounts for the spatial entanglement below the critical temperature for condensation ($T_C$).

\section{The relationship between spatial entanglement and ODLRO}
\begin{figure}[t]\centering
{\includegraphics[width=0.3\textwidth]{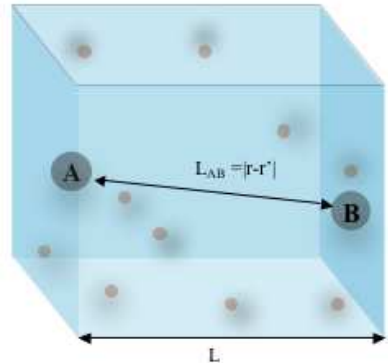}
\caption{A confining box of volume $V=L^3$ contains a Bose gas (small spheres) at temperature $T$.  Two two-level systems $A$ and $B$ (large darker spheres) interact with two localised regions of the gas, $\Omega_A$ and $\Omega_B$  separated by $L_{AB}=|\vec{r}-\vec{r}^{\, \prime}|$.  }\label{system}}
\end{figure}

Let us first discuss the system in question.  We take a gas of non-interacting Bosons in thermal equilibrium described by the temperature density operator 
\begin{equation}
\rho_G=\frac{1}{Z}e^{-\beta(\hat{H}_G-\mu\hat{N})},
\end{equation}
of the grand canonical ensemble, where $\hat{H}_G$ is the Hamiltonian of the gas,  $\beta=1/k_BT$ where $k_B$ is Boltzmann's constant and $T$ is the temperature of the system, $Z$ is the grand partition function and $\hat{N}$ being the number operator for all Bosons. The chemical potential $\mu$  that accounts for the particle number fluctuations is fixed for any temperature by the implicit relation $N=\sum_{\vec{k}}1/(e^{\beta(E_{\vec{k}}-\mu)}-1)$ .

The gas is contained in a quantisation box of volume $V=L^3$ (Fig.~\ref{system}) with periodic boundary conditions and the Bosons occupy energy eigenmodes, weighted by the temperature, characterised by the plane waves $\phi_{\vec{k}}(\vec{r})=V^{-\frac{1}{2}}\exp(i\vec{k}\cdot\vec{r})$, where $k_i=2\pi l_i/L_i$, $i=x,y,z$ and $l_i$ are the wave quantum numbers in the three spatial dimensions. The free Hamiltonian of the non-interacting gas is $\hat{H}_G=\sum_{\vec k} \, E_{\vec k} \, \hat{a}_{\vec k}^{\dagger} \, \hat{a}_{\vec k}$, where $\hat{a}_{\vec k}^\dagger=\int_Vd\vec{r}\,\phi_{\vec{k}}(\vec{r})\Psi^{\dagger}(\vec{r})$ and $\hat{a}_{\vec{k}}$ are the creation and annihilation operators of Bosons in the ${\vec k}$-th mode with energies $E_{\vec k} = {\hbar^2 {\vec{k}}^2 \over 2m}$.

\subsection{Interaction of gas with localised probes}\label{sec:interaction}

As we are interested in entanglement between distant parts of the gas we need to define two separate regions ($A$ and $B$).   This can be achieved in a very natural way using the effective volumes of two probe systems  $\Omega_A$ and $\Omega_B$, which are approximated by spheres of radius $R$, see (Fig.~\ref{system}).  

Here, the probes are two general two-level systems that come into contact with the regions $A$ and $B$ for some short interaction time $t_{int}$.  Probe $S_A$ interacts only with region $A$ and probe $S_B$ interacts only with region $B$.  The probes start in the unexcited state $\rho_{P} = |00\rangle\langle00|$ (where $|00\rangle=|0\rangle_A|0\rangle_B$) and are therefore initially separable w.r.t. one another.  Therefore, as their interaction with the gas is local, any entanglement that is found between the probes afterwards must have come from the particle number entanglement in the gas, as local operations alone cannot create entanglement.  The interaction Hamiltonian 
\begin{equation}
H_I=Q_A\sigma^{+}_{S_A} + Q_B\sigma^+_{S_B}
\end{equation}
says that if there are some quanta inside a region (detected by the operator $Q_A$ for region $A$), the probe interacting with that region becomes excited (i.e. $\sigma^+_{S_A} |0\rangle_A = |1\rangle_A$).  A realistic way to implement this interaction is discussed in section \ref{sec:resultsanddiscussion}. 
 
The full Hamiltonian for the time-evolution consists of the free evolution of the gas, $H_G$ and the interaction between the gas and the probes weighted with some coupling $\Gamma(t)$ which may be time-dependent, $H(t)=H_G + \Gamma(t)H_I$.

The initial state of the probes and the gas is $\rho(0)=\rho_G\otimes\rho_P$ and if the interaction time $t_{int}=\delta t$ is short enough one can express the unitary time evolution as
\begin{equation}
 U=\exp\bigg{(}i \int_0^{\delta t}  \d t \, H(t)\bigg{)}\approx U (\delta t) \approx[1+i(\delta t H_G+\Gamma H_I)]
 \end{equation}
  where $\Gamma=\int_0^{\delta t} \d t \, \Gamma(t)$.  After the interaction the probes and the gas are described by the un-normalised state 
$\rho(\delta t)=U (\delta t) \, \rho(0) \,U^{\dagger} (\delta t) $,
which  can be reduced to the state of the probes, $\rho_{AB}$, by tracing out the gas degrees of freedom.  If we consider the limit of very small interaction time $\delta t <<1$ but finite, $ \Gamma <<1$  we find the probes in the state
\begin{eqnarray} \label{eq:rhoAB}
	\rho_{AB} &\propto& \tr_G[
		(1 + i \Gamma ( Q_A \, \sigma^+_{S_A} + Q_B \, \sigma^+_{S_B}) ) \, 
		\rho_G  \nonumber\\
		& &\otimes |00\> \< 00| \,
		(1 - i \Gamma ( Q_A \, \sigma^+_{S_A} + Q_B \, \sigma^+_{S_B}) ) ] \\
	 		&\propto&  |00\> \< 00|
		+ \Gamma^2 \,  \tr_G[ 
		(Q_A \, |10\>  + Q_B \, |01\> )\dots
		\nonumber\\
		&&\times  \, \rho_G  \,
		( Q_A \, \< 10| +  Q_B \, \< 01| )] \nonumber\\
			&& + i \Gamma \, \tr_G[
		 (Q_A \, |10\>\< 00|  +  Q_B \, |01\>\< 00| )  \, \rho_G ] 
		 + \mbox{ h.c.}.\nonumber
\end{eqnarray}
After rearranging and normalising we obtain the final state of the probes after the interaction written in the basis $|00\>, |10\>, |01\>, |11\>$,
\begin{equation} \label{eq:probesAB}
	\rho_{AB}=\frac{1}{N}
		\left[\begin{array}{cccc}
		1 						& i \Gamma \<  Q_A \> 	
		& i \Gamma \<  Q_B \> 			& 0	\\
		- i \Gamma \<  Q_A \> 	& \Gamma^2 \<  Q_A^2 \> 
		& \Gamma^2 \<  Q_A Q_B \> 	& 0\\
		- i \Gamma \<  Q_B \> 	& \Gamma^2 \<  Q_A  Q_B \>
		& \Gamma^2 \<  Q_B^2 \> 	& 0  \\
	 	0 		& 0		& 0		& 0
	\end{array}\right],
\end{equation}
where $\langle \hat O\rangle$ denote the expectation values over the gas, $\tr [\hat O \, \rho_G]$. %
Here the normalisation is $N=1+\Gamma^2(\langle Q_A^2\rangle +\langle Q_B^2\rangle)$. Note, that for short interaction times the probability of the probes both receiving a kick, represented by the right-hand bottom corner of the matrix , the $|11\rangle\langle11|$ element, representing $\rho_{AB}$, scales as $\Gamma^4$ which is much smaller than the other terms and will be neglected. This is an approximation for short interaction times, which we shall speak about again later.

The position observables are 
\begin{equation}
Q_{A(B)}=\int_{\Omega_{A(B)}} \d \vec{r} \, \Psi^{\dagger} (\vec{r}) \, f_{A(B)}(\vec{r}) \, \Psi(\vec{r}),
\end{equation}
where $f_{A(B)}(\vec{r})$ is a real, positive function with support only in $A(B)$ that defines the shape of our regions.  Here we will take $f_{A(B)}(\vec{r})=C$ to be equal to a constant inside the region and zero outside.  The coefficients $\langle Q_A^2\rangle$ etc. are evaluated in the number basis and are
\begin{eqnarray}
\langle Q_{A(B)}\rangle &=& \int \d \vec{r} \, f_{A(B)}(\vec{r}) \, \rho_1(\vec{r},\vec{r}),\nonumber\\
\langle Q^2_{A(B)}\rangle& =& \int \d \vec{r} \, f^2_{A(B)}(\vec{r}) \, \rho_1(\vec{r},\vec{r}) +\nonumber\\
& &n^2 \, \int \d \vec{r} \, \d \vec{r}^{\, \prime} \, f_{A(B)}(\vec{r}) \, f_{A(B)}(\vec{r}^{\, \prime}) \, g(\vec{r},\vec{r}^{\, \prime};\vec{r},\vec{r}^{\, \prime}),\nonumber\\
\langle Q_A \, Q_B\rangle &=& n^2 \, \int \d \vec{r} \, \d \vec{r}^{\, \prime} \, f_A(\vec{r}) \, f_B(\vec{r'})\, g(\vec{r},\vec{r}^{\, \prime};\vec{r},\vec{r}^{\, \prime}).
\end{eqnarray}
where $n$ is the density of Bosons.

The state of the probes has been related to the reduced one-body density operator $\rho_1(\vec{r},\vec{r}^{\, \prime})=\langle\Psi^{\dagger}(\vec{r})\Psi(\vec{r}^{\, \prime})\rangle$ and the pair distribution function 
\begin{equation}
g(\vec{r},\vec{r}^{\, \prime};\vec{r},\vec{r}^{\, \prime})=n^{-2} \langle \Psi^{\dagger}(\vec{r})\Psi^\dagger(\vec{r}^{\, \prime})\Psi(\vec{r})\Psi(\vec{r}^{\, \prime})\rangle
\end{equation}
 of the gas. As the gas is translationally invariant one can express $\rho_1$ and $g$ as functions of only the relative distances, $\vec{r}-\vec{r}^{\, \prime}$, which leads to
\begin{eqnarray}
\label{onebodydensity}
		\rho_1(\vec{r} - \vec{r}^{\, \prime})
	&=& {1 \over V} \, \sum_{\vec k} \, 
	\langle \hat{a}_{\vec k}^{\dagger} \, \hat{a}_{\vec k} \rangle \,
	e^{i \vec{k} \cdot (\vec{r}-\vec{r}^{\, \prime})},\\
	g(\vec{r} - \vec{r}^{\, \prime})
	&=& {1 \over \langle N\rangle^{2}} \,
	\sum_{\vec p, \vec q, \vec k} \, 
	\langle \hat{a}_{\vec p}^{\dagger} \, \hat{a}_{\vec q}^{\dagger} \, 
	\hat{a}_{\vec q-\vec k} \, \hat{a}_{\vec p+\vec k} \rangle \,
	e^{i \vec{k} \cdot(\vec{r}-\vec{r}^{\, \prime})},
\end{eqnarray}
 where we have used the expresssion $\Psi^{\dagger}(\vec{r})=\sum_{\vec{k}}\phi_{\vec{k}}(\vec{r})\hat{a}^{\dagger}_{\vec{k}}$.  Additionally for a non-interacting Bose gas one can then write the pair distribution function as the square of the one-body density matrix $g(\vec{r}-\vec{r}^{\, \prime})=1+(\rho_1(\vec{r}-\vec{r}^{\, \prime})/n)^2$.

\subsection{Entanglement of the probes}\label{sec:entofprobes}

Let us take a closer look at the final state of the probes $\rho_{AB}$ (\ref{eq:probesAB}). 
The presence of the $|00\rangle\langle00|$ accounts for the `non-events' where neither of the probes have interacted with the gas.  This massive `background' has to be subtracted to reveal any effects of an interaction and possibly entanglement arising from it.  If we include this term in the analysis of entanglement of course we find that the probes are in a separable state w.r.t. one another.  However if we look at the sub-ensemble where the probes interacted with the system,  we find entanglement.  Mathematically we project $\rho_{AB}$ into the subspace spanned by the projector $\hat{P}=\mathbf{1}-|00\rangle\langle00|$ to the new density matrix $\rho_{AB}'$.  The two probes are now in an entangled state, but because we have used a global operation, i.e. non-local operation, we can therefore not simply claim that we have {\it extracted} any entanglement from the gas, we could have produced it by our projection.  Nevertheless, as the same projection has been applied to the probes regardless of the temperature of the gas, any temperature dependence of the entanglement of the probes comes from a change of the spatial entanglement with temperature inside the gas. The maximum amount of `false' entanglement that has been generated by this projection will be calculated later and will be taken as our zero (background) level.

We can now investigate the increase in spatial entanglement of the gas from the false background level by quantifying the entanglement between probes.  By using the strategy of inserting two-level probes into the gas that pick up the entanglement we have managed to reduce a many-body entanglement problem into a typical qubit problem and hence we can apply the negativity \cite{Eisert01,Vidal02}, a standard entanglement measure, to quantify the entanglement. The negativity is defined as the sum of the negative eigenvalues of a partially transposed density matrix \cite{Peres96}.

  The partial transposed matrix of $\rho_{AB}^{\prime}$,
 \begin{equation}
	\rho^{\prime \, T_B}_{AB} = \frac{1}{\<  Q_A^2 \> + \<  Q_B^2 \>}
		\left[\begin{array}{cccc}
		0			& 0 				& 0		& \<  Q_A Q_B \>\\
		0			& \<  Q_A^2 \>  	& 0 		& 0	\\
		0 			& 0				& \<  Q_B^2 \> 		& 0  \\
	 	\<  Q_A Q_B \> 	& 0		& 0		& 0
	\end{array}\right],
\end{equation}
reveals  one negative eigenvalue and the negativity $\mathcal{N}$, that measures the amount of entanglement in $\rho_{AB}'$ is then
\begin{eqnarray}
\label{negativity}
\mathcal{N}&=& \frac{\langle Q_A Q_B\rangle}{\langle Q_A^2\rangle +\langle Q_B^2\rangle}\\
&=& \frac{1}{2} \frac{n^2\Omega^2 + \int_{\Omega_A} \d \vec{r} \, \int_{\Omega_B} \d \vec{r}^{\, \prime} \, \rho_1(\vec{r}-\vec{r}^{\, \prime})^2}{n\Omega + n^2\Omega^2 +\int_{\Omega_A} \d \vec{r} \, \int_{\Omega_A} \d \vec{r}^{\, \prime} \, \rho_1(\vec{r}-\vec{r}^{\, \prime})^2}\nonumber
\end{eqnarray}
where we have taken the volumes of the two regions to be equal, $\Omega_A=\Omega_B=\Omega$ and where $f_{A(B)}$ are taken as top-hat functions within the respective volumes and zero outside.   

It is clear to see that for a fixed density $n$ and probe volume $\Omega$, the negativity is affected by the behaviour of the reduced one-body density matrix $\rho_1(\vec{r}-\vec{r}^{\, \prime})$ alone. We note again that the presence of ODLRO in a BEC depends entirely on the behaviour of the reduced one-body density matrix $\rho_1(\vec{r}-\vec{r}^{\, \prime})$ at large separations.   Thus the amount of spatial entanglement is directly related to the existence of ODLRO.  To find out how precisely the nature of this dependence is we proceed to calculate the negativity explicitly.

\subsection{Calculation of the negativity}\label{sec:negativity}

The presence of ODLRO in a Bose gas is indicative of the new condensed phase so our task now is to compute the integrals in the negativity $\mathcal{N}$ and confirm that there is an increase in the amount of entanglement below $T_C$ as (\ref{negativity}) suggests. There are two kinds of double integrals in $\mathcal{N}$, namely, a cross-over term for the interaction between regions $A$ and $B$
\begin{equation}
\label{int1}
\mathcal{I}_1^{AB}=\int_{\Omega_A} \d \vec{r} \, \int_{\Omega_B} \d \vec{r}^{\, \prime}\, \rho_1(\vec{r}-\vec{r}^{\, \prime})^2
\end{equation}
 and a on-site term, for instance in the volume $\Omega_A$
 \begin{equation}
 \mathcal{I}_1^{AA}=\int_{\Omega_A} \d \vec{r}\, \int_{\Omega_A} \d \vec{r}^{\, \prime} \, \rho_1(\vec{r}-\vec{r}^{\, \prime})^2.
 \end{equation}
For evaluation we use the expression given in (\ref{onebodydensity}) as our starting point and remember that we must treat the ground state separately when temperatures below $T_C$ are considered.  We approximate the sum over momenta in equation (\ref{onebodydensity}) as an integral and evaluate it for low $k$ as for temperatures around $T_C$ high momenta modes have a infinitesimal occupation.  We can then write the one-body density operator as
\begin{equation}
\label{onebody2}
\rho_1(\vec{r}-\vec{r}^{\, \prime})=\frac{z}{\lambda^2}\frac{\exp\big{(}-\sqrt{4\pi (1-z)}|\vec{r}-\vec{r}^{\, \prime}|)/\lambda\big{)}}{|\vec{r}-\vec{r}^{\, \prime}|} \Big{(}+n_0\Big{)},
\end{equation}
where $n_0$ is the ground state density and should only be included when the gas is below $T_C$ \cite{Pitaevskii03}.  The fugacity is denoted by $z=\exp(\mu/k_BT)$ and $\lambda$ is the thermal wavelength given by $\lambda=\sqrt{h^2/2\pi m k_B T}=(2.612/n)^{\frac{1}{3}}(T_C/T)^{\frac{1}{2}}$, where in the last equality we have used an expression for $T_C$ \cite{Pitaevskii03}.  

We now evaluate the two integrals above $T_C$ where the ground state occupation number $n_0$ can be  neglected.  For spherical volumes of the probes, $\Omega=\frac{4\pi}{3}R^3$, we obtain after some calculation $\mathcal{I}_1^{AA}$ to be
\begin{eqnarray}
\mathcal{I}_1^{AA}(T>T_C) = & &\frac{2\pi^2z^2}{\kappa^4\lambda^4}\Big{[}1-2\kappa^2R^2+\frac{8}{3}\kappa^3R^3\nonumber\\
& &-(1+2\kappa R)\exp\big{(}-2\kappa R\big{)}\Big{]},
\end{eqnarray}
 where $\kappa=2\sqrt{4\pi(1-z)}/\lambda$.
For $\mathcal{I}_1^{AB}$, the two regions $A$ and $B$ are far apart and their radius $R$ shall be much smaller than the distance between them. Thus the distance $|\vec{r}-\vec{r}^{\, \prime}|$ in the integral $\mathcal{I}_1^{AB}$ with $\vec{r}$ in $\Omega_A$ and $\vec{r}'$ in $\Omega_B$ or {\it vice versa} is large and can be set to a constant $L_{AB}\approx|\vec{r}-\vec{r}^{\, \prime}|$ as it does not change much over the comparatively small volume of the individual regions $\Omega$. As we are investigating the relationship of spatial entanglement to ODLRO, the distance $L_{AB}$ between the two regions is taken to infinity $L_{AB}\rightarrow \infty$, which means that the reduced one-body density matrix in the cross-over integral (\ref{int1}) vanishes which means the integral is also zero,  $\mathcal{I}_1^{AB}(T>T_C)=0$.
 
Below $T_C$ we must include the ground state density $n_0$ becomes macroscopic and must be included in $\rho_1(\vec{r}-\vec{r}^{\, \prime})$ (\ref{onebody2}.  This results in a non-zero cross-over term, $\mathcal{I}_1^{AB} (T<T_C)= n_0^2\Omega^2$, where again the distance $L_{AB}\rightarrow\infty$.  The on-site terms is comprised of three contributions that arise from the inclusion of the ground state density in the one-body reduced density operator $\rho_1(\vec{r}-\vec{r}')$.  The full integral  $\mathcal{I}_1^{AA}(T<T_C)$ was found to be
\begin{eqnarray}
\label{rho1aa}
\mathcal{I}_1^{AA}(T<T_C)=n_0^2\Omega^2+\mathcal{I}_1^{AA}(T>T_C)+n_0\mathcal{I}',
\end{eqnarray}
where 
\begin{equation}
\mathcal{I}'=\frac{128\pi^2 z}{\kappa^5\lambda^2}[4-\kappa^2R^2+\frac{1}{3}\kappa^3R^3
-(4+4\kappa R +\kappa^2 R^2)e^{-\kappa R}].
\end{equation}
%
Here the ground state density is $n_0=n[1-(T/T_C)^{\frac{3}{2}}]$. We set the ground state energy to zero  and use $\mu = -k_B T \ln(1+1/N_0)$  \cite{Pitaevskii03} to determine the small correction to the chemical potential.   

\subsection{Results and Discussion}\label{sec:resultsanddiscussion}

With the integrals computed the focus can shift back to the negativity (\ref{negativity}), which should be evaluated for a range of temperatures.  However, care must be taken to chose the parameters carefully as the short interaction time approximation, which enabled us to set the coefficient of the $|11\rangle\langle11|$ to zero, no longer holds for some values of $\Gamma n \Omega$.  In $\rho_{AB}$ (\ref{eq:probesAB}), the terms that contain the highest powers of  $\Gamma n\Omega$ are $\Gamma^2\langle Q_A^2\rangle\propto \Gamma^2n^2\Omega^2$ that corresponds to the states $|01\rangle\langle01|$ and $|10\rangle\langle10|$   and the matrix element corresponding to $|11\rangle\langle11|$ has terms with the powers $\Gamma^4\langle Q_AQ_BQ_AQ_B\rangle\propto\Gamma^4n^4\Omega^4$.  Thus for our approximation to be valid the quantity $\Gamma n \Omega<<1$ must stay small so that it is still safe to neglect the $|11\rangle\langle11|$ term from the final state of the probes.  In contrast if $\Gamma n\Omega>1$ even though $\Gamma<<1$, the state $|11\rangle\langle11|$ would participate.

We would now like to see how the negativity alters in the neighbourhood of $T_C$.  To correctly plot the expression in (\ref{negativity}) we need to weight it with the probability that the probes have interacted with the gas $\textrm{tr}_{p}[\hat{P}\rho_{AB}\hat{P}^{\dagger}]$.  Only if the interaction happened will the probes be in the state $\rho_{AB}'$ with negativity $\mathcal{N}$. The probability $\tr_{p} [\hat{P} \, \rho_{AB}\,\hat{P}^{\dagger}]$ for the probes to have interacted with the gas, with $\langle Q_A^2\rangle=\langle Q_B^2\rangle$, is
\begin{eqnarray}
\tr_{p} [\hat{P} \, \rho_{AB}\,\hat{P}^{\dagger}]&=&\tr_{p}\big{[}({\bf1}-|00\rangle\langle00|)\rho_{AB}({\bf{1}}-|00\rangle\langle00|)\big{]}\nonumber\\
&=&\frac{\Gamma^2}{N}\tr_p\big{[}\langle Q_A^2\rangle|10\rangle\langle10|+\langle Q_B^2\rangle|01\rangle\langle01|\dots\nonumber\\
& &+(\langle Q_AQ_B\rangle|10\rangle\langle01|+\textrm{h.c.})\big{]}\nonumber\\
&=&\frac{2\Gamma^2\langle Q_A^2\rangle}{1+2\Gamma^2\langle Q_A^2\rangle}.
\end{eqnarray}
It is important that the negativity is weighted in this way as there is no point talking about entanglement between the probes if they have never interacted with the gas.   The probability is also temperature dependent and the resulting quantity is the averaged or weighted entanglement $\mathcal{E}$ given by

\begin{figure}[t]\centering
{\includegraphics[width=0.5\textwidth]{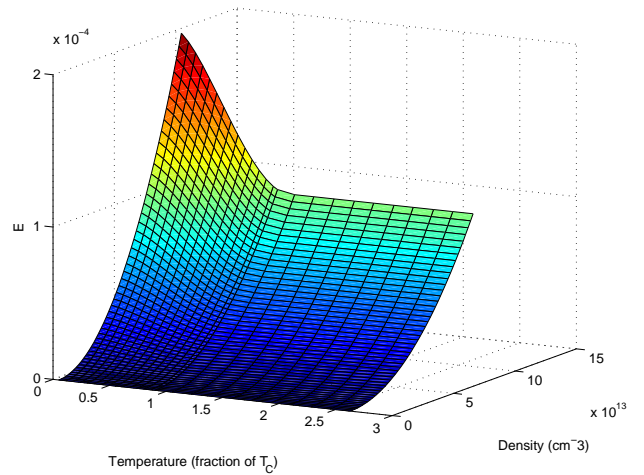}
\caption{The surface depicts the amount of weighted entanglement $\mathcal{E}$  (\ref{extractableentanglement}) for $\Gamma=2.4\times10^{-5}$, $R=10^{-4}$cm and $N=10^6$ as a function of the temperature of the gas and the particle density.     Above $T_C$ ($T_C=1$ on our scale), the constant amount of entanglement between the probes is exactly the maximum amount of false entanglement $\mathcal{E}_F$ (see text). As the temperature drops below $T_C$, the amount of entanglement between the probes increases significantly above the background level $\mathcal{E}_F$.  This shows that spatial entanglement is directly related to the onset of ODLRO  below $T_C$. }\label{fig:ExEnt}}
\end{figure}

\begin{eqnarray}
\label{extractableentanglement}
\mathcal{E}&=& \tr_{p} [\hat{P} \, \rho_{AB}\,\hat{P}^{\dagger}]\, \mathcal{N}\nonumber\\
&=&\frac{\Gamma^2(n^2\Omega^2+\mathcal{I}_1^{AB})}{1+2\Gamma^2(n\Omega+n^2\Omega^2+\mathcal{I}_1^{AA})},
\end{eqnarray}
The weighted entanglement is the average entanglement between an ensemble of probes where only a certain fraction of the probes, given by the probability  $\tr_{p} [\hat{P} \, \rho_{AB}\,\hat{P}^{\dagger}]$, have interacted with the gas.   If one could know for certain that two probes interacted with the gas then the entanglement between them would be given by the negativity (\ref{negativity}), but such entanglement measures are defined on ensembles.
The plot of $\mathcal{E}$ is shown in Fig. \ref{fig:ExEnt}.  

The actual amount of entanglement picked up by the probes is extremely small compared to a maximally entangled state which would have a negativity of one half.  This is because we have calculated the weighted entanglement and there is a low probability for the probes to interact with the gas in the short interaction time.   If the probes always interacted with the gas, so that the contribution in $\rho_{AB}$ (\ref{eq:probesAB}) of the $|00\rangle\langle00|$ term were zero, we would end up with a  maximally entangled state if $n\Omega>>1$ at very low temperatures, $T\rightarrow 0$.  We can see this by looking at the negativity (\ref{negativity}) and noting that the integrals $\mathcal{I}_1^{AA}$ and $\mathcal{I}'$ go to zero in these limits.  The terms with $n^2\Omega^2$ would then dominate and the negativity would tend to one half, $\mathcal{N}\rightarrow1/2$.  However this is not the case in this paper as  $\Gamma n \Omega<<1$ because we have chosen a very short interaction time parametrised by $\Gamma$ and very small probe volumes $\Omega\approx10^{-12}$cm$^{3}$.  Although note that we have a typical condensate density of $n=10^{14}$cm$^{-3}$. 

Although the entanglement between the probes is small, the amount of weighted entanglement $\mathcal{E}$ behaves as expected.  It is constant above the critical temperature and only sharply increases below the condensation temperature $T_C$ where ODLRO is present (see Fig.~\ref{fig:ExEnt}). 
So indeed, the entanglement between two distant regions $A$ and $B$ in the Bose gas, which was picked up by the probes, is directly related to the existence of ODLRO in the gas, which emerges below $T_C$.  The constant amount of entanglement above $T_C$ is artificial and results from the global - non-local - operation we made that post-selected only those scattering events when the gas and probes interacted and got rid of the background of non-events.

We should therefore calculate the amount of false entanglement $\mathcal{E}_{F}$ between the probes due to the projection and compare this with the constant amount of entanglement above $T_C$.   Consider a general, pure and symmetric product of the two probes state $|\Psi\rangle=|\psi\rangle_A\otimes|\psi\rangle_B$ where $|\psi\rangle$ is parametrised by $\epsilon$, $|\psi\rangle = \sqrt{1-\epsilon}|0\rangle+\sqrt{\epsilon}|1\rangle$.  If the probes were in this state then they would be separable and the negativity would vanish.  However we can create entanglement by acting on $|\Psi\rangle$ globally with the projector $\hat{P}=\mathbf{1}-|00\rangle\langle00|$.  To represent the short interaction time approximation, $\epsilon<<1$ is taken and the term where the two probes are both excited, $|11\rangle$, can be set to zero.  The amount of false entanglement generated is $\mathcal{E}_F=\epsilon^2(1-\epsilon)^2/(1-\epsilon^2)$, where here the negativity has been weighted with the probability of finding the probes in the state spanned by the projector $\hat{P}$.  If we choose $\epsilon=0.01$ so that it is consistent with $\Gamma n \Omega \approx 0.01$ used in our calculation, the amount of false entanglement generated by the non-local projection on $|\Psi\rangle$ is $\mathcal{E}_F\approx10^{-4}$.  The amount of false entanglement  calculated here is identical to the constant amount of entanglement above $T_C$ for the real state $\rho'_{AB}$ shown in Fig.~\ref{fig:ExEnt},   namely  $\mathcal{E}=1.09 \cdot 10^{-4}$ for $n=10^{14}\textrm{cm}^{-3}$ . 

This demonstrates that above $T_C$ the entanglement between the probes occurs solely because of the global projection, but below $T_C$ the increase in entanglement between the probes exceeds $\mathcal{E}_F$ and must come from the entanglement of the gas.
Interestingly we could use $\mathcal{E}_F$ as the reference level of entanglement, set it to zero and use a suitably scaled $\mathcal{E}$ as an order parameter for a BEC, which must be zero above $T_C$ and finite below $T_C$. 
 \\\\
We will now discuss the relationship between our work and another paper \cite{Ferrer00} that also investigates the interaction of two probe-type particles with a Bosonic field.  Firstly we note that the first term of the one body reduced density matrix (\ref{onebody2}) is a Yukawa-type interaction potential and the second term is the ground state density.  We have seen that as the distance between the probes goes to infinity the Yukawa interaction term goes to zero and only the ground state density is left.  It is the filling of the ground state that is responsible for the ODLRO that is also responsible for the spatial entanglement.

In fact two massive particles interacting via a Bosonic field is discussed in \cite{Ferrer00} and the authors found that below the critical temperature for condensation there is an infinite ranged interaction between the particles, which was attributed to the collective occupation of all the Bosons in the ground state.Ê
We have seen that it is the macroscopic occupation of theÊground state that is responsible for the spatial entanglement in the gas and causes two separable probes that interact locally with the gas to become entangled. By reformulating the problem in terms of second quantised spatial modes we have shown that two non-local properties of a BEC, spatial entanglement and infinite-ranged forces are equivalent.Ê

Let us finish this section with a brief discussion of how one may implement the probe-gas interaction experimentally.  We need to think of a physical scattering process between probe and the gas  where a number of bosons in a region interacts with an internal degree of freedom of the probe.  This type of interaction could be realised by inserting two probe atoms into the gas that are much heavier than the surrounding Bosons.  During the short interaction time the probes would not move about much but an internal degree of freedom, for instance the nucleus spin, may become excited when hit by a Boson. 

\section{Conclusion}\label{sec:conclusions}

In this paper we have presented an operational method of investigating entanglement between spatial modes in a BEC.  Spatial entanglement exists between two distant regions below the critical temperature for condensation due to the massive increase in population of the ground state energy level.  This is evident in the negativity, Eq.~(\ref{negativity}), due to its dependence on the reduced one-body density matrix.   As ODLRO is also defined through the behaviour of the one-body reduced density matrix and exists only below $T_C$, from the results in this paper we can conclude that ODLRO is responsible for the spatial entanglement of the gas.  

\medskip

VV and LH acknowledge the support of the Engineering and Physical Sciences Research Council in UK for funding and the National University of Singapore for their hospitality.  JA acknowledges support of the Gottlieb Daimler und Karl Benz-Stiftung.  DK would like to thank M. Wiesniak for  interesting discussions. All authors thank B.-G. Englert for helpful comments.


\begin{thebibliography}{99}
\bibitem{LRO} The simplest form of ODLRO that is found in a BEC is also known as long-range order. see for instance \cite{Yang62, Pitaevskii03}.
\end{thebibliography}
\end{document}